\documentclass[aps,pre,preprint,superscriptaddress]{revtex4-1}

\def\la{\langle}
\def\ra{\rangle}
\def\rv{{\bf r}}
\def\qv{{\bf q}}

\def\kB{k_{B}}
\def\qs{q^{*}}
\def\q0{q_{0}^{*}}
\def\Rg0{R_{g0}}
\def\chie{\chi_{e}}
\def\chia{\chi^{*}_{a}}
\def\Nbar{\overline{N}}
\def\chibar{\overline{\chi}_{e}}

\usepackage{graphicx,amsmath,xcolor}
\begin{document}


\title{Test of a scaling hypothesis for the structure factor of
       disordered diblock copolymer melts}


\author{Jens Glaser}
\affiliation{Department of Chemical Engineering and Materials Science,
University of Minnesota, Washington Ave SE 421, Minneapolis 55455, MN, USA}


\author{Jian Qin}
\affiliation{Department of Chemical Engineering and Materials Science,
University of Minnesota, Washington Ave SE 421, Minneapolis 55455, MN, USA}
\affiliation{Department of Chemical Engineering, Pennsylvania State University, University Park, State College, 16803, PA, USA}
\author{Pavani Medapuram}
\affiliation{Department of Chemical Engineering and Materials Science,
University of Minnesota, Washington Ave SE 421, Minneapolis 55455, MN, USA}
\author{Marcus M\"uller}
\affiliation{Institut f\"ur Theoretische Physik, Georg-August Universit\"at, Friedrich-Hund-Platz 1, 37077 G\"ottingen, Germany}
\author{David Morse}
\email[Corresponding author: $\ $]{morse012@umn.edu}
\affiliation{Department of Chemical Engineering and Materials Science,
University of Minnesota, Washington Ave SE 421, Minneapolis 55455, MN, USA}


\date{\today}

\begin{abstract}
Coarse-grained theories of dense polymer liquids such as block copolymer melts predict a universal dependence of equilibrium properties on a few dimensionless parameters. For symmetric diblock copolymer melts, such theories predict a universal dependence on only $\chi N$ and $\Nbar$, where $\chi$ is an effective interaction parameter, $N$ is a degree of polymerization, and $\Nbar$ is a measure of overlap.  We test whether simulation results for the structure factor $S(q)$ obtained from several different simulation models are consistent with this two-parameter scaling hypothesis. We compare results from three models: (1) a lattice Monte Carlo model, the bond-fluctuation model, (2) a bead-spring model with harsh repulsive interactions, similar to that of Kremer and Grest, and (3) a bead-spring model with very soft repulsion between beads, and strongly overlapping beads. We compare results from pairs of simulations of different models that have been designed to have matched values of $\Nbar$, over a range of values of $\chi N$ and $N$, and devise methods to test the scaling hypothesis without relying on any prediction for how the phenomenological interaction parameter $\chi$ depends on more microscopic parameters. The results strongly support the scaling hypothesis, even for rather short chains, confirming that it is indeed possible to give an accurate universal description of simulation models that differ in many details.
\end{abstract}


\maketitle

\section{Introduction}
Modern understanding of polymer liquids relies heavily on the study of highly simplified models, and of theories that predict highly universal behavior. The best developed example of this is the theory of dilute and semidilute polymer solutions in good solvent \cite{DeGennes1979,DesCloizeaux1990,Schafer1999}. Experiments and simulations in this regime are well described by a scaling theory that predicts a universal dependence on only two dimensionless parameters, which measure the extent of overlap and the strength of the bare repulsion per chain. The theory is based on an analysis of a highly simplified model, due to Edwards \cite{Edwards1965}, of polymers as continuous random walks with a short-ranged repulsion between segments. Perhaps the most important step in testing this theory was the demonstration that experimental data for polymer solutions is consistent with the scaling hypothesis, which was achieved by comparing data from a variety of chemical systems, over a range of conditions and chain lengths \cite{Noda1984,Takahashi1985}.

We are interested here in dense polymer liquids, such as block copolymer melts and polymer blends, rather than solutions.  Coarse-grained theories of these systems are based on a simple generalization of the Edwards model. To describe dense liquids, the model is studied in a nearly-incompressible limit in which density fluctuations are strongly suppressed.  Following Matsen, we will refer to this as the ``standard model'' \cite{Matsen2002} for this class of liquids. The purpose of the current paper is to verify a scaling hypothesis suggested by a long series \cite{Fredrickson1987,Holyst1993,Kudlay2003,Wang2002a,Grzywacz2007,Beckrich2007,Morse2006,Morse2009,Morse2011,Qin2012} of theoretical studies on this model, which predict behavior that depends only on the parameters required as inputs to self-consistent field theory and one additional parameter, $\bar{N}$, which is a measure of overlap. Specifically, we test a prediction that the structure factor $S(q)$ in disordered symmetric diblock copolymer melts should exhibit a universal dependence on two parameters $\chi N$ and $\Nbar$, independent of many details of different simulation models and experimental systems. This is tested by comparing simulation results for $S(q)$ obtained from several different simulation models over a range of chain lengths and parameter values.  

\section{Theoretical Background}
The vast majority of theoretical work on block copolymers and polymer mixtures has thus far been based on the self-consistent field (SCF) approximation \cite{Leibler1980,Matsen1994,Matsen1995,Tyler2005}. Self-consistent field theory (SCFT) is a coarse-grained density functional theory that expresses the free energy functional of an inhomogeneous polymer liquid as a sum of the free energy of an inhomogeneous ideal gas of random-walk polymers and an excess free energy, and in which the excess free energy is taken to be a local functional of composition. In the simplest and most common form of the theory, for systems with two types of monomers A and B, the excess free energy per monomer $f_{\rm ex}$ is taken to be a function $f_{\rm ex} = kT \chi(T) \phi_{A}\phi_{B}$, in which $\phi_{i}$ represents the local average volume fraction of $i$ monomers, and $\chi$ is the Flory-Huggins interaction parameter. 

SCF theory predicts a universal dependence of all equilibrium properties on a small set of dimensionless parameters. In systems with two types of monomer, such as diblock copolymer melts and binary polymer blends, SCF predictions all depend on a product $\chi N$ that quantifies the excess free energy of mixing per chain, where $N$ is a degree of polymerization. Other relevant parameters include ratios of chain and block lengths, macroscopic composition (in polymer mixtures), and ratios of statistical segment lengths or radii of gyrations. Thus, for example, for idealized diblock copolymer melts with monomers of equal statistical segment lengths $b=b_{A}=b_{B}$, SCFT predicts a universal phase diagram \cite{Leibler1980,Matsen1994} that depends on only $\chi N$ and the volume fraction $f_{A}$ of the $A$ block.   

We focus in this contribution on the behavior of the structure factor $S(q)$ in simulations of symmetric diblock copolymer melts. $S(q)$ is a correlation function for composition fluctuations, defined by
\begin{equation}
        S(q) = \int \! d\rv \; \la \psi(\rv)\psi(0) \ra e^{i\qv\cdot \rv} \quad,
\end{equation}
where $\psi(\rv) = [\delta c_{A}(\rv) - \delta c_{B}(\rv)]/2$ is a composition field.  Here, $\delta c_{i}(\rv)$ is a deviation of the number concentration of $i$ monomers from the spatial average value $c_i = c/2$ ($c$: total average monomer concentration). 

It is possible to use the SCF approximation to calculate $S(q)$, by employing the relationship between correlation functions and functional derivatives of $F$. This yields the so-called random phase approximation (RPA) \cite{Leibler1980}. The RPA yields a prediction for the quantity $cNS^{-1}(q)$ in symmetric diblock copolymer melts of the form
\begin{equation}
   cN S^{-1}(q) = F(q \Rg0) - 2 \chie N
\label{eq:sqrpa}
\end{equation}
where $F(x)$ is a known analytic function, $\Rg0 = \sqrt{Nb^{2}/6}$ is the unperturbed radius of gyration, and $\chie$ is an effective interaction parameter. In the most general form of SCFT, the effective interaction parameter that appears in the RPA is given by a derivative $\chi_e \equiv (2 k_B T)^{-1}\partial^{2} f^{ex}(\phi_{A})/\partial \phi_{A}^{2}$ of the SCF excess free energy per monomer. In the special case of symmetric diblock copolymer melts, the RPA thus predicts that $S(q)$ depends on one non-dimensionalized wavenumber $q\Rg0$ and a single thermodynamic parameter $\chie N$. 

In any theory that is based on the standard model, physical predictions must be invariant under changes in how we define "monomer". The standard model is based upon a description of polymer conformations as idealized fractals (Wiener processes), with no chemical microstructure.  As such, the definition of a monomer, or the number N of monomers per chain, is arbitrary. Consequently, all SCF and RPA predictions can be expressed as relationships among dimensionless quantities that are invariant in this sense. Among such invariants are the product $\chie N$, which is a measure of excess free energy per chain for transferring an A chain of $N$ monomers into a B-rich environment, the invariant scattering intensity $S(q)/(Nc)$, and quantities such as $q \Rg0$ in which a length scale is non-dimensionalized by comparing it to a measure of the coil size. This invariance principle also, however, admits one parameter that does not appear in SCFT predictions (except as a trivial prefactor in expressions for total free energy), which is a measure of dimensionless concentration.  In an incompressible one-component liquid of chains of length $N$, with a monomer concentration $c$ and a polymer concentration $c/N$, we may define an invariant concentration 
\begin{equation}
   \bar{C} \equiv cR^{3}_{e0}/N = N^{1/2} cb^{3}
\end{equation} 
which is the number of chains in a volume $R^{3}_{e0}$, where $R _{e0} \equiv \sqrt{N}b$ is a random-walk approximation for the root-mean-squared end-to-end length. Alternatively, following Fredrickson and Helfand \cite{Fredrickson1987}, we may define an ``invariant degree of polymerization''
\begin{equation}
    \Nbar \equiv \bar{C}^{2} \equiv N(cb^{3})^{2}
    \quad. \label{eq:barn}
\end{equation}
We use $R_{e0}$ to define $\bar{C}$ and $\Nbar$, rather than the radius of gyration or some other measure of coil size, simply for consistency with the definition of $\Nbar$ used in previous work \cite{Fredrickson1987, Morse2011,Qin2012}. This choice is arbitrary. Notably, SCF predictions depend non-trivially upon all of the dimensionless parameters that are allowed by the invariance principle {\it except} the parameter $\bar{C}$ or $\Nbar$.
  
The predicted absence of any dependence on $\Nbar$ is a peculiarity of the SCF approximation. A series of closely related one-loop theories \cite{Fredrickson1987, Grzywacz2007,Beckrich2007,Morse2011,Qin2012} of the standard model, which go beyond the SCF approximation, all yield predictions that depend upon $\Nbar$ as well as the parameters that appear in SCF predictions. Moreover, these theories all predict results that reduce to corresponding SCF or RPA predictions in the limit $\Nbar \rightarrow \infty$. These studies thus suggest that the SCF approximation is exact in the limit of infinitely strongly overlapping chains, but not finite chains.  For symmetric diblock copolymer melts, the renormalized one-loop (ROL) theory predicts a structure factor that is given by a universal function
\begin{equation}
  cN S^{-1}(q) = H(q \Rg0, \chie N, \Nbar)
\label{eq:sq}
\end{equation}
of $q\Rg0$, $\chie N$, and $\Nbar$, where $\Rg0 = \sqrt{N/6}b$. The renormalized one-loop theory \cite{Grzywacz2007,Morse2011,Qin2012} yields an explicit prediction in which the function $H$ is given as the sum of an RPA prediction (the RHS of Eq. (\ref{eq:sqrpa}) with renormalized values of $\chie$ and $b$, plus a correction that is proportional to $\Nbar^{-1/2}$.  One-loop theories of fluctuations predict ${\cal O}(\Nbar^{-1/2})$ corrections to SCF predictions appear in many other quantities, including the critical region for polymer blends and corrections to the random walk statistics. It appears that this one-loop correction is simply the first term in a renormalized loop expansion of corrections to RPA and SCF predictions that (if pursued further) would yield a Taylor expansion of $H(q\Rg0, \chie N, \Nbar)$ in powers of $\Nbar^{-1/2}$, in which the ROL theory for $S(q)$ retains only the first two terms \cite{Morse2011}.  We thus expect the existence of a scaling relation of the form given in Eq. (\ref{eq:sq}) to have a wider range of validity than the one-loop approximation.

Extensive simulations of diblock copolymer melts have been carried out previously using a variety of coarse-grained lattice and continuum models. The use of coarse-grained models for this purpose presumes a high degree of universality, both among different simulation models and between coarse-grained simulations and experiments.  Because none of these coarse-grained models provide an accurate representation of any real polymer liquid, it would be pointless to study them unless one believes that the equivalent behavior can be obtained from corresponding states of models or physical systems that differ in many details. The goal of the present contribution is to test the extent to which data for $S(q)$ from several different simulation models of diblock copolymers is consistent with the existence of a universal function of the form given in Eq. (\ref{eq:sq}). 

The scaling hypothesis of Eq. (\ref{eq:sq}) simply postulates that the invariant scattering intensity $S(q)/cN$ exhibits a universal dependence upon all of the invariant dimensionless parameters that appear in the standard model. As such, a test of the adequacy of Eq. (\ref{eq:sq}) is also a test of the adequacy of the standard model as a generic description of models and experimental systems that differ in many microscopic details.

\section{Simulation methodology}
\label{sec:methods}
We have studied three coarse-grained models. The first model is a lattice bond fluctuation model,  which has been described in Refs.~\cite{Carmesin1988,Mueller1995,Mueller1995a}.  The second is a bead-spring model with a steeply repulsive Weeks-Chandler-Anderson (WCA) pair repulsion, similar to that used by Kremer and Grest \cite{Kremer1990}. The third is a bead-spring model with a much softer pair interaction that allows monomers to overlap \cite{Groot1998,Pike2009,Wang2009a}. We will refer to these models in what follows as models L (``lattice"), H (``hard"), and S (``soft"), respectively.

Model L (the bond fluctuation model) is based on a cubic lattice, with a lattice spacing $\sigma$, in which each monomer occupies (i.e., excludes other monomers from) the eight lattice sites of a cube of volume $\sigma^3$. The bond between consecutive monomers along a chain is allowed to have any of $108$ vectors with lengths up to $\sqrt{10}\sigma$, excluding the vector of length $\sqrt{8}\sigma$ in the $(2,2,0)$ direction to avoid bond crossing. Contacts between monomers of type $i$ and $j$ are assigned an energy cost $\epsilon_{ij}$, with $\epsilon_{AA}=\epsilon_{BB}=-\epsilon_{AB} = \alpha/2$. Simulations of this model used a monomer concentration $c \sigma^{3} = 1/16$, corresponding to half of the maximum allowed filling. Model L was simulated using lattice Monte Carlo, using a combination of local moves and of chain flip moves that interchange the A and B blocks of a chain.

Model H used here is identical to that used in Ref. \cite{Qin2012}. In this model, all monomers interact via a purely repulsive Weeks-Chandler-Anderson (WCA) \cite{Weeks1971} pair potential, for which 
\begin{equation}
    V_{ij}(r) = 4\epsilon_{ij} [(\sigma/x)^{12}-(\sigma/x)^{6} + 1/4]
\end{equation}
for $i,j=A,B$, with $V_{ij}(r) = 0$ beyond a cut-off $r_c=2^{1/6} \sigma$, with $\epsilon_{AA} = \epsilon_{BB} = \kB T$ and $\epsilon_{AB} = \epsilon_{AA} + \alpha$. Bonded monomers interact via a harmonic potential $v_{\rm bond}( r ) = \kappa (r - l)^2/2$, with $\kappa = 400 \kB T/ \sigma^{2}$ and $l = \sigma$. All simulations of this model were carried out in NVT ensemble in a $L \times L \times L$ periodic cubic simulation cell, using a monomer concentration $c = 0.7 \sigma^{-3}$. Data for this model was shown previously in Ref. \cite{Qin2012}. The model was simulated using a replica-exchange hybrid MC algorithm, in which short MD runs are used as proposed MC moves, among other MC moves, as described in Ref. \cite{Qin2012}.

Model S uses a soft repulsive interaction
\begin{equation} 
  V_{ij}(r) =\epsilon_{ij} [(r/\sigma)-1]^2
\end{equation}
cut off at $r_c=\sigma$. This was originally introduced by Groot and coworkers \cite{Groot1998} in DPD simulations. We do not use a DPD thermostat here, instead we use a Nos\'e-Hoover thermostat.  Bonded particles in this model experience an additional harmonic bond potential with zero rest length and a spring constant $\kappa = 3.406 k_BT/\sigma^2$.  Simulations of the model used a monomer concentration $c = 3.0 \sigma^{-3}$, more than four times that of model H, thus creating substantial overlap between monomers. Model S was also simulated using a replica exchange hybrid MC algorithm (Simpatico \footnote{See https://github.com/dmorse/simpatico}), using hybrid MC and chain flips moves, in which the HOOMD-blue code \cite{Anderson2008} code was used as an engine to accelerate the hybrid MD moves on graphics processing units.

In each of these three models, the pair interaction between monomers of types $i$ and $j$ is proportional to a parameter $\epsilon_{ij}$, with $\epsilon_{AA} = \epsilon_{BB}$. For each model, we define a parameter $\alpha = \epsilon_{AB} - \epsilon_{AA}$, which we use as a control parameter to vary $\chie$. 

The structure factor in a diblock copolymer melt exhibits a peak at a wavenumber $q^{*}$ with a peak intensity $S(q^{*})$ that increases with increasing degree of thermodynamic repulsion between $A$ and $B$ monomers, or increasing $\chie$. An example of this behavior, from our simulations of model S, is shown in Fig. 1. In experiment, $\chie$ can be controlled by varying temperature. In the simulations presented here, we vary $\chie$ by varying the control parameter $\alpha$ , while keeping the temperature and all other parameters of the Hamiltonian fixed. For each model and chain length, we have conducted simulations over a range of $\alpha \geq 0$, up to values near those at which an ordered phase forms.  We have determined values for the
peak wavenumber $q^{*}$ and peak intensity $S(q^{*})$ for each simulation by fitting $S(q^{*})$ to a smooth function of $q$. An example of the quality of the fit is shown in Fig. 1. 

\begin{figure}
\centering\includegraphics[width=\columnwidth]{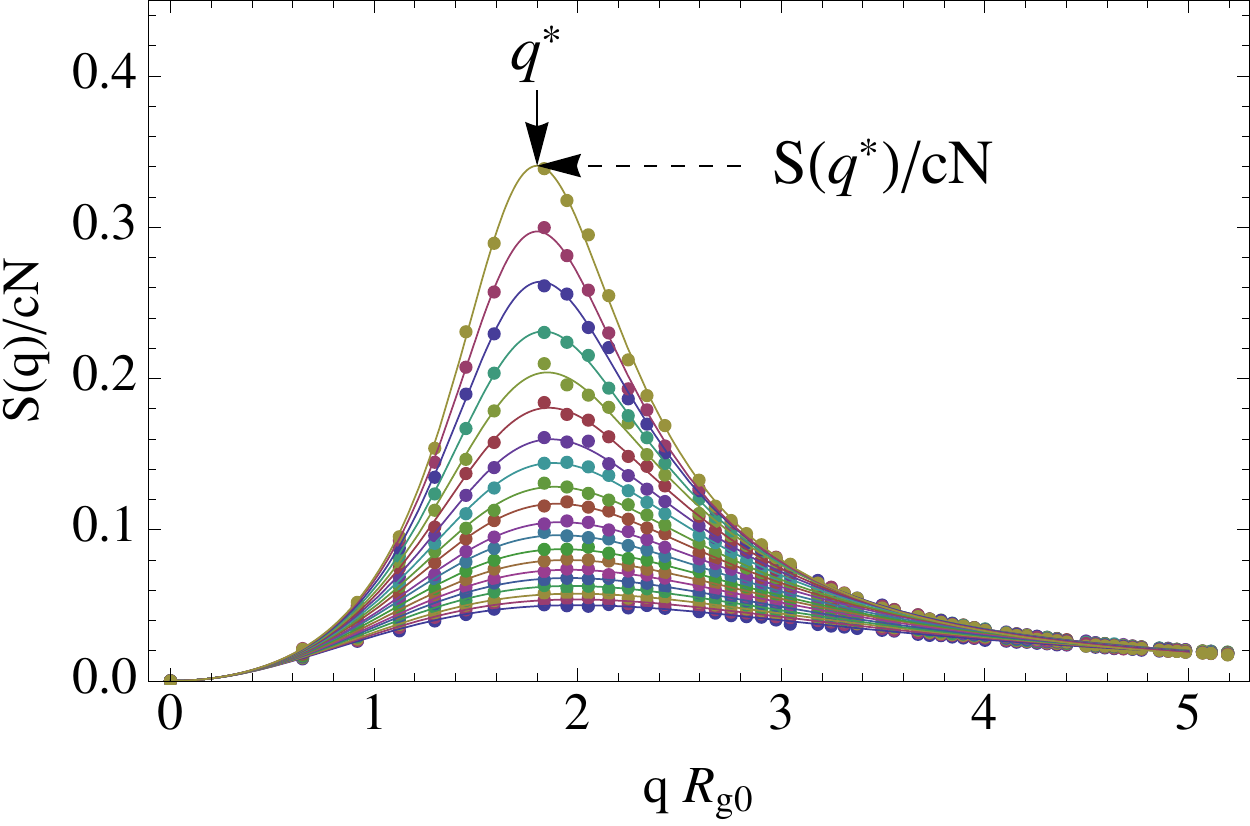}
\caption{Reduced structure factor $S(q)/cN$ as a function of the dimensionless wave number $q \Rg0$ of a symmetric diblock copolymer melt for model S for $N=32$ ($\bar N=480$), for a range of values $\alpha=\epsilon_{AB}-\epsilon_{AA} = 0, \dots, 1.35$ (from bottom to top).}
\label{fig:sq}
\end{figure}

Our analysis requires a value of the statistical segment length $b$, which is used to define the parameters $\Rg0 = \sqrt{N/6}b$ and $\Nbar \equiv N(c b^{3})^{2}$. A value of $b$ was determined for each model by the procedure used in Refs. \cite{Morse2009,Qin2012} and in earlier work \cite{Wittmer2007} by Wittmer and coworkers. For each model, we measure an $N$-dependent apparent segment length $b^{2}(N) \equiv 6 R_{g}^{2}/N$  in a homopolymer reference state (with $\alpha = 0$), for several different values of the chain length $N$. The chain length dependence of $b(N)$ is very well described by the renormalized one-loop theory for mono-disperse homopolymer melts.  We define the parameter $b$ as the limit $b^{2} = \lim_{N \rightarrow \infty} b^{2}(N)$. This definition is required for consistency with the renormalized one-loop theory, which predicts random walk behavior only in this limit.  This yields values of $b = 3.244 a$ for model L, $b = 1.404\sigma$ for model H and $b = 1.088 \sigma$ for model S. 
 
The scaling hypothesis predicts that we should obtain equivalent behavior from different simulation models, over a range of values of $\alpha$, from simulations of systems with equal values of $\Nbar$.  To test this, we have designed pairs of simulations of different models in which the chain lengths have been adjusted to give matched values of $\Nbar = N(cb^{3})^2$. Because different models have different values of the dimensionless parameter $cb^{3}$, this requires that we compare simulations of different models with different values of $N$, the number of monomers per chain. To facilitate the design of simulations of models H and S with matched values of $\Nbar$, the value of the spring constant parameter $\kappa$ used for model S was chosen so as to yield a value of $(c b^{3})^{2} = 14.9$ that is exactly four times the value of this parameter obtained for model H. Simulations of model S with chains of $N$ monomers thus have the same value of $\Nbar$ as that obtained in simulations of model H using chains with $4N$ monomers. Below, we thus compare simulations of model S with chains of length $N=16$ and $N=32$ to model H simulations of much longer chains of length $N=64$ and $N=128$. 

\section{Testing Universality}
To test the scaling hypothesis of Eq.~\eqref{eq:sq}, we compare results for the non-dimensional peak wavenumber $q^{*}\Rg0$ and invariant peak intensity $cNS^{-1}(q^{*})$ for the three simulation models described above. Eq. \eqref{eq:sq} implies that these quantities should both be universal functions 
\begin{eqnarray}
   q^{*}\Rg0       & = & Q^{*}(\chie N, \Nbar)  \label{qstarScale} \\
   cNS^{-1}(q^{*}) & = & H^{*}(\chie N, \Nbar)  \label{SstarScale}
\end{eqnarray}
of the two variables $\chie N$ and $\Nbar$, where $\Rg0 = \sqrt{N/6}b$.  The RPA predicts a constant value of $q^{*}\Rg0 = 1.95$, independent of $\chie  N$, and a value for $H^{*}$ that varies linearly with $\chie N$, independent of $\Nbar$. As part of our hypothesis, we interpret the microscopic parameters $\chie$ and $b$ for each model as ``renormalized'' parameters that can exhibit an arbitrary dependence on temperature, monomer, concentration and on the details of the monomer-scale interactions, but that are assumed to be independent of chain length. Specifically, we assume that the value of $\chie$ for each of our simulation models is some unknown, generally nonlinear function $\chie(\alpha)$ of the control parameter $\alpha$, and that the function $\chie(\alpha)$ is different for different models. 

The scaling hypothesis predicts that we should obtain equivalent behavior in corresponding states of different models. Corresponding states of symmetric diblock copolymer melts  are defined as states with equal values of the $\Nbar$ and of the product $\chi_{e}N$. A useful alternative definition corresponding states may be obtained by defining an invariant interaction parameter $\chibar$, by requiring that $\chibar \Nbar = \chi_{e}N$. This definition yields
\begin{equation}
    \chibar \equiv \chi_{e}/(cb^{3})^2 \quad.
\end{equation}
Corresponding states can then also be described as states with equal value of $\chibar$ and $\Nbar$, which is equivalent to requiring equal values of $\chibar \Nbar$ and $\Nbar$. We assume that $\chibar$, like $\chie$, is a model-dependent function of $\alpha$, but is independent of $N$. This definition is needed in subsection \ref{subsec:mapping}, where we define corresponding values of $\alpha$ in different models by requiring that they yield equal values for $\chibar$.

In what follows, we always compare results of pairs of simulations of different models in which the values of $N$ have been chosen to give equal of $\Nbar$.  After creating simulations with matched values of $\Nbar$, the remaining conceptual difficulty that must be overcome to test Eqs.~\eqref{qstarScale} and \eqref{SstarScale} is the the fact that the equations postulate a dependence on a parameter $\chie(\alpha)$ that is an unknown function of the control parameter $\alpha$, and that cannot be directly measured. Two strategies are used in what follows to overcome this difficulty. The first, which is described in subsection A, is to test the universality of the relationship between two measurable quantities, the peak wavenumber and peak intensity. The second, which is described in subection C, involves an attempt to collapse the data from different models, over a range of chain lengths, by constructing an explicit mapping between corresponding values of $\alpha$ in different models. 

\subsection{Peak wavenumber vs. peak intensity}
\label{sec:qstarchia}
As our first test of universality, we construct parametric plots of the relationship between observables $q^{*}\Rg0$ and $cN S^{-1}(q^{*})$ for pairs of models with matched values of $\Nbar$, over a range of values of the control parameter $\alpha$ for each model. Eq.~\eqref{SstarScale} implies that systems with equal values of both $\Nbar$ and $cN S^{-1}(q^{*})$ must be in corresponding states, with equal values of $\chie N$. Systems with equal values of $\bar{N}$ and $cN S^{-1}(q^{*})$ should thus also exhibit equal values for the peak wavenumber $q^{*}\Rg0$. This implies that a parametric plot of $q^\star\Rg0$ vs. $cN S^{-1}(q^{*})$ should yield a curve whose form depends on $\Nbar$, but that data from different simulation models with identical values of $\Nbar$ should collapse onto the same master curve.

\begin{figure}
\centering\includegraphics[width=\columnwidth]{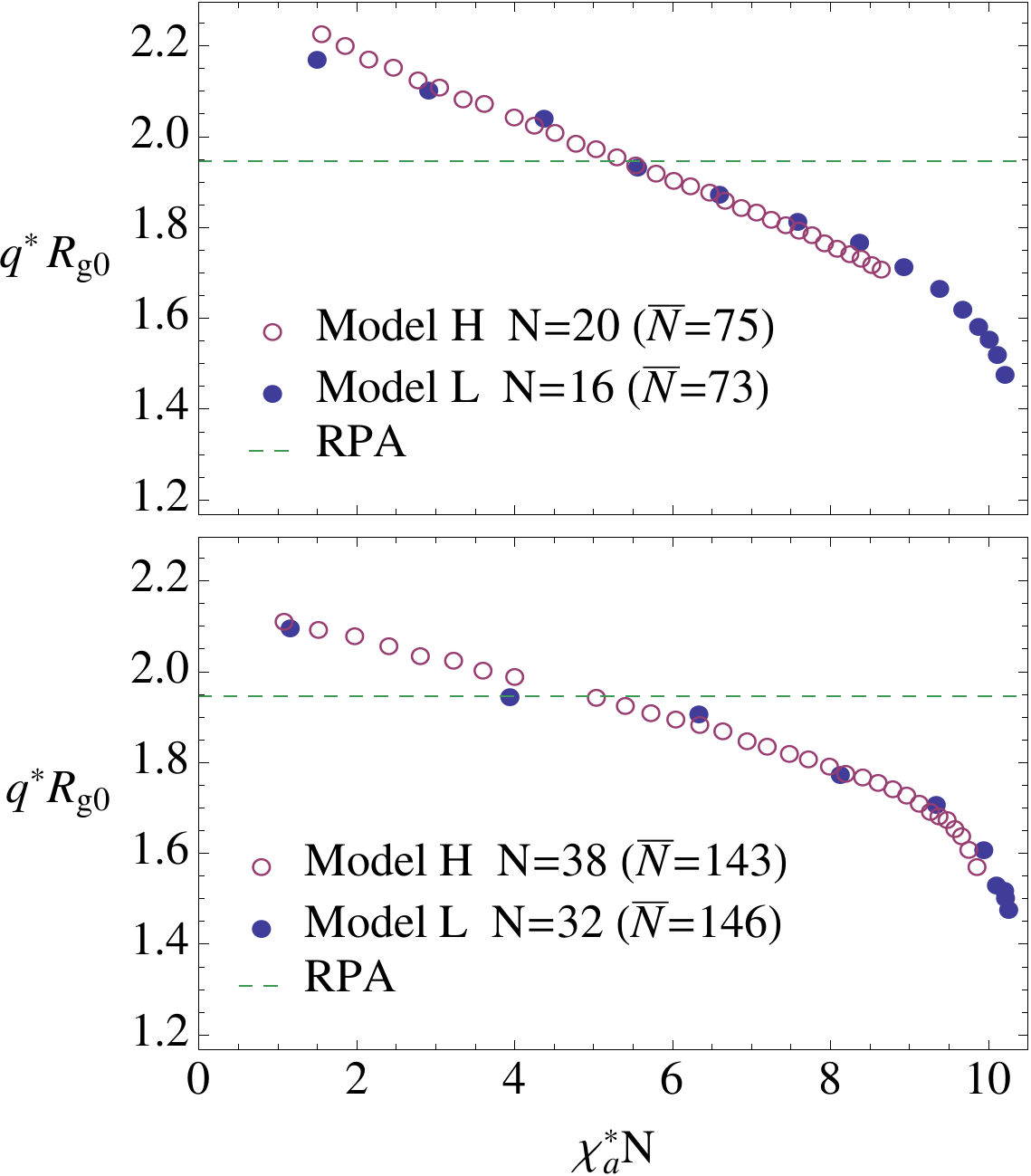}
\caption{Nondimensional peak scattering location $\qs R_{g0}$, vs. apparent interaction parameter $\chi_a^* N$ for simulations of model L (the lattice bond fluctuation model) and model H (a bead spring model with harsh repulsion). Data is shown for two pairs of simulations with nearly matched invariant degrees of polymerization $\Nbar \simeq74$ (upper panel) and $\Nbar \simeq 145$ (lower panel). The RPA prediction $q^{*}R_{g0}=1.95$ is shown as a horizontal dashed line.}
\label{fig:wca_bfm}
\end{figure}

\begin{figure}
\centering\includegraphics[width=\columnwidth]{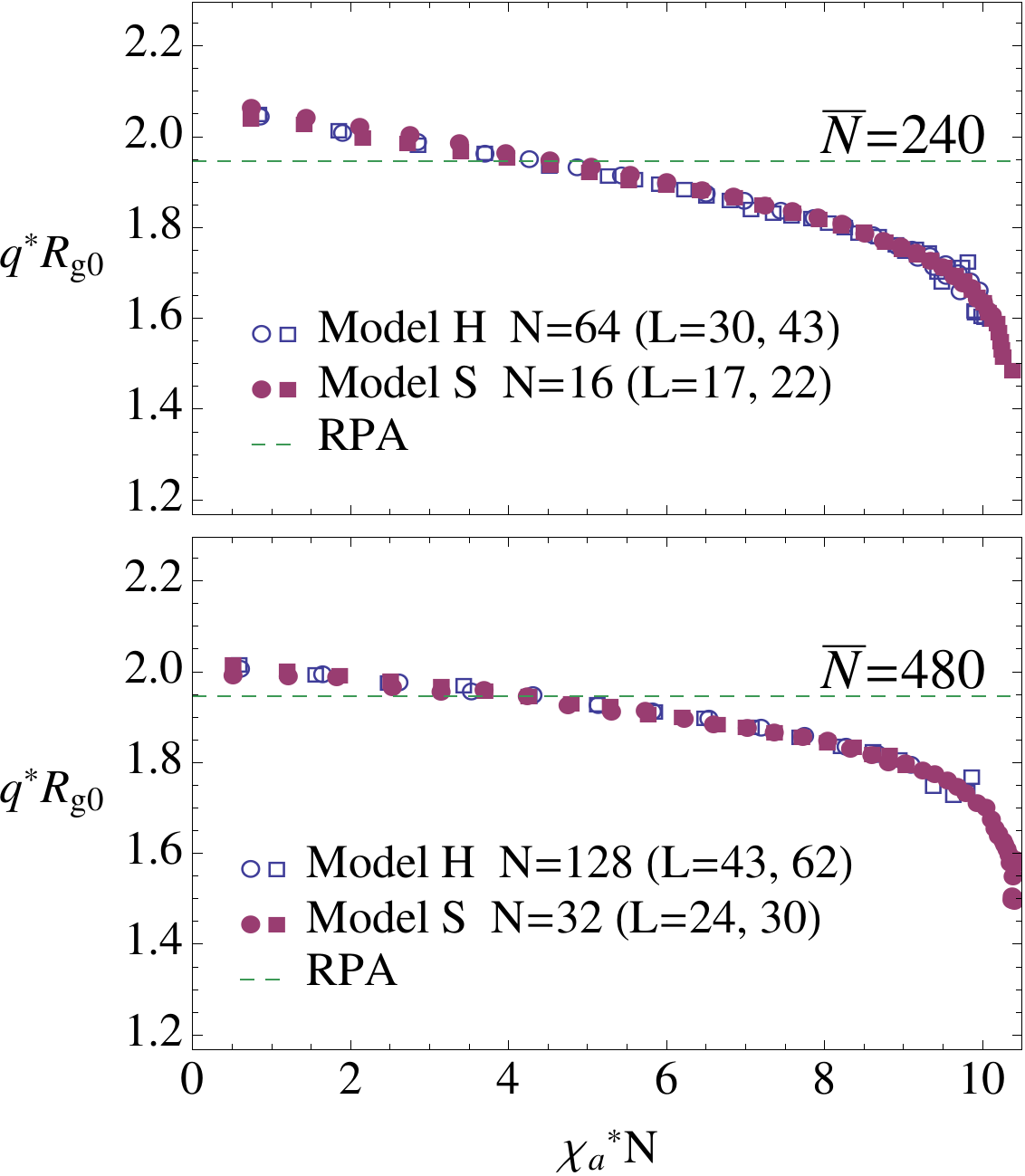}
\caption{Normalized peak scattering location $\qs R_{g0}$ vs. $\chi_a^*N$ 
for model H (open symbols) and model S (filled symbols). Data is shown for 
two pairs of simulations with matched values of $\Nbar = 240$ (upper panel) 
and $\Nbar = 480$ (lower panel). For each model, data is shown for two 
different values of the simulation box dimension $L$, using circles for 
the smaller box and squares for the larger, as indicated in the legend. 
The RPA prediction is shown by a dashed horizontal line. Data for model 
$H$ are shown over ranges $\alpha=$ 0 - 1.2 for $N=64$ and 
$\alpha=$ 0 - 0.4 for $N=128$. Data for model model S are shown over
ranges $\alpha=$ 0 - 6.4  for $N=16$ and $\alpha =$ 0 - 4.7 for $N=32$.}
\label{fig:wcadpd}
\end{figure}

Figures \ref{fig:wca_bfm} and \ref{fig:wcadpd} show plots of $q^{*}\Rg0$ vs. a parameter $\chia N$ that is directly related to the scattering intensity, for several pairs of simulations of different models with matched values of $\Nbar$.  The quantity $\chia$ is an ``apparent'' interaction parameter, also used in Ref. \cite{Qin2012}, that we define to be the value of $\chi$ that would be inferred by fitting the measured peak intensity to the RPA prediction, by requiring that
\begin{equation}
     c N S^{-1}(q^{*}) \equiv 2 [ (\chi  N)_{c} - \chia N ]
\end{equation}
where $(\chi  N)_c \equiv F(q^{*} \Rg0)/2 = 10.495$ \cite{Leibler1980} is the critical value of $\chie  N$ for symmetric diblock copolymers, at which the RPA predicts a divergence in $S(q^{*})$. Plotting $q\Rg0$ vs. $\chia N$ is thus equivalent to plotting of $q\Rg0$ vs. $cNS^{-1}(q^{*})$, but show the data in a form in which peak intensity increases from left to right, and in which $\chia N = 10.495$ corresponds to a divergence of $S(q^{*})$. The RPA also predicts a constant value of $q^{*}\Rg0 = 1.95$ that depends on neither $\chi_a^* N$ nor $\Nbar$, which is shown by a horizontal dashed line in these figures.

Figure \ref{fig:wca_bfm} shows simulation results for $\qs \Rg0$
vs. $\chi_a^* N$ for models L and H, for pairs of simulations in which the
values of $N$ have been chosen to give nearly matched values of
$\Nbar$ (to the closest integer value of $N$). The upper panel shows a
comparison of simulations with $\Nbar \simeq 75$ and the lower panel
shows a pair of simulations with $\Nbar \simeq 145$.  When we compare
data from simulations using matched values of $\bar{N}$, the level of
agreement between the lattice and continuum model is remarkable.


Figure \ref{fig:wcadpd} shows data for $\qs/q_0$ vs. $\chi_a^* N$ for the H (circles) and S (triangles) models, for two pairs of simulations with exactly matched values $\Nbar$ = 240 (upper panel) and $\Nbar$=480 (lower panel). In each of these pairs of simulations, matched values of $\Nbar$ are obtained by using chains in model H that have 4 times as many monomers as those used in the corresponding model S, because of the much higher monomer concentration in the latter model. Here, data for each model is shown for two different values of the simulation box length $L$, to verify the absence of finite size effects. Agreement between the different models is again nearly perfect. This collapse of the data for models with such widely disparate chain lengths provides a particularly clear confirmation that the form of this function depends only on overlap parameter $\Nbar$, and not on the actual number of monomers.

\subsection{Failure of Perturbation Theory}
\label{sec:failure_perturb}

\begin{figure}
\centering\includegraphics[width=\columnwidth]{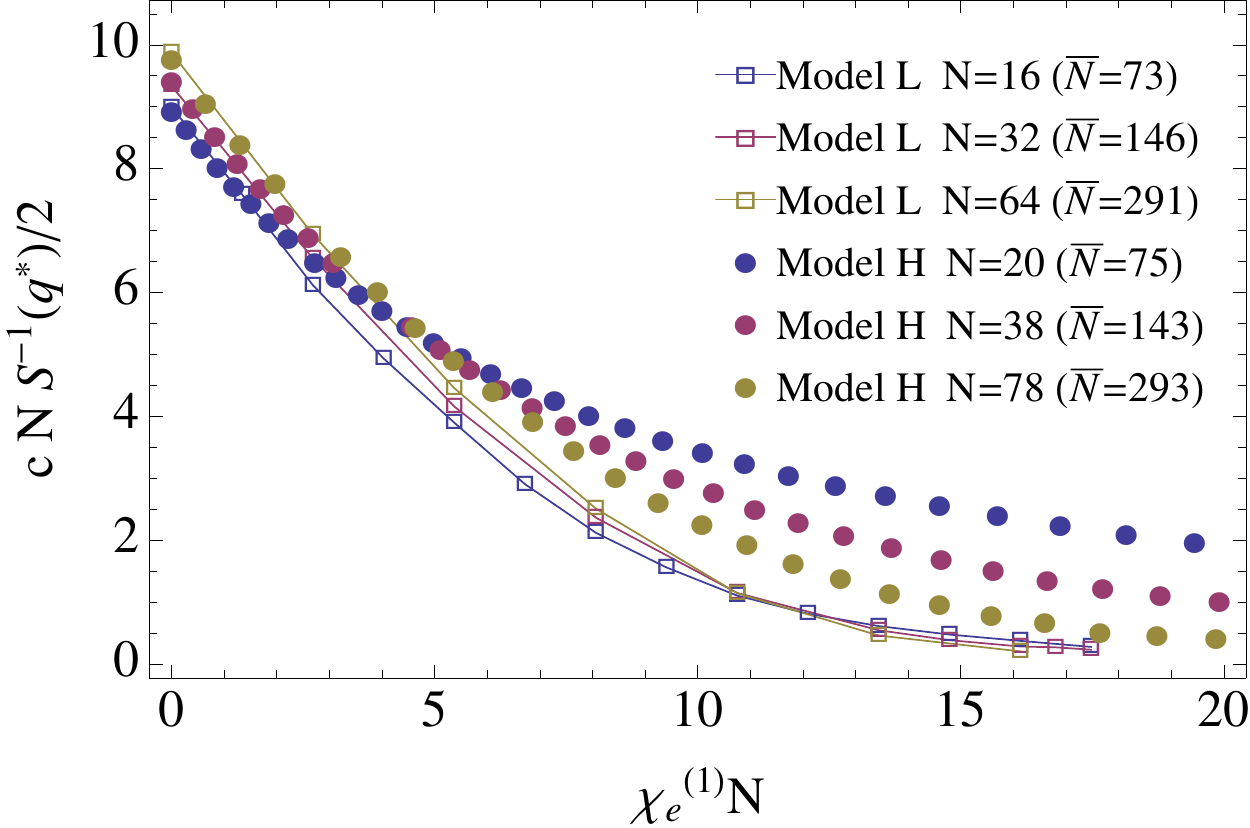}
\caption{Normalized inverse peak intensity $(1/2) cN S^{-1}(q*)$ for the lattice bond fluctuation (L) model (lines and open squares) and the continuum 'hard' (H) model (solid circles), for three pairs of simulations with nearly matched values of $\Nbar =$ 75, 145 and 292. Data for both models is plotted here vs. $\chi_e^{(1)}N=z^{\infty}\alpha N/\kB T$, where $z^{\infty}_{\mathrm{H}} = 0.2965$ for model H and
$z^{\infty}_{\mathrm{L}}=4.2$ for model L.}
\label{fig:nomap_wca_bfm}
\end{figure}

In previous work \cite{Qin2012}, we compared results for the invariant peak intensity for model H to several different theories by using perturbation theory to estimate $\chi_{e}(\alpha)$. In Ref.~ \cite{Morse2009}, it was shown how $\chie(\alpha)$ could be calculated to first-order in a Taylor expansion in $\alpha$, of the form
\begin{equation}
   \chie (\alpha) = z^\infty \alpha/k_B T + \mathcal{O}(\alpha^2)
\end{equation}
in which the value of the coefficient $z^{\infty}$, known as the effective coordination number, can be obtained from simulations of a reference homopolymer ($\alpha = 0$) melt. This analysis generalized and provided a more rigorous statement of an earlier proposal by one of us (M.M.) and Kurt Binder \cite{Mueller1995,Muller1999} that $\chie$ be estimated for the lattice model by taking $\chie(\alpha) = z(N) \alpha/k_B T$, where $z(N)$ was the number of inter-molecular neighbors of each site in a lattice simulation of a reference homopolymer melt containing chains of length $N$. The corresponding definition of $z(N)$ for a continuum model is also a measure of the number of inter-molecular neighbors of each monomer, which is expressed in this case as an integral involving the inter-molecular radial distribution function. Ref. \cite{Morse2009} clarified that, in detailed comparisons with theory, the SCF parameter $\chie$ must be defined using a value $z^{\infty}$ that is defined by extrapolating $z(N)$ to $N=\infty$. Following the prescription outlined in that reference, we have obtained values of $z^{\infty}= 4.2$ for the lattice model, $z^{\infty} = 0.2965$ for model H, and $z^{\infty} = 0.237$ for model S. 

Comparisons of data for $c N S^{-1}(q^{*})$ vs. $\chie N$ to the (ROL) theory, using this estimate for $\chie$, showed good agreement at low values of $\alpha$, but exhibited systematic deviations at higher values of $\alpha$ \cite{Qin2012}. Because large deviations occurred only at large values of $\alpha$, they were tentatively ascribed to a breakdown in the first-order expansion for $\chie(\alpha)$. In that study, however, we could not cleanly separate failures of the Taylor expansion of $\chie(\alpha)$ from failures of the ROL theory.

A more basic test of the adequacy of a first-order perturbation theory for $\chie(\alpha)$ may be obtained by using this estimate for $\chie(\alpha)$ to compare two different models, using pairs of simulations with matched values of $\bar{N}$. Fig. \ref{fig:nomap_wca_bfm} shows a comparison of data for $cN S^{-1}(q^{*})$ vs. $\chie(\alpha)$ for three pairs of simulations of models L and H, plotting using the estimate $z^{\infty}\alpha N/\kB T$, for values of $\bar{N} \simeq $ 75, 145 and 245. When plotted in this way, the results from the two models disagree rather badly, and agree only at low values of $\chie N$.  This indicates a severe failure of either the scaling hypothesis or of the first-order approximation for $\chie(\alpha)$. In the next subsection, we show that it is the latter.

\subsection{Mapping between corresponding states}
\label{subsec:mapping}
To test Eq. (\ref{SstarScale}) without making any assumptions about how $\chie$ depends on $\alpha$, we have tested whether it is possible to collapse the data for different models by constructing an explicit nonlinear mapping between corresponding values of the control parameter $\alpha$ in different models. 

Consider a comparison of two models, 1 and 2, with control parameters $\alpha_{1}$ and $\alpha_{2}$. We assume that $\chibar$ is given by different functions of $\alpha$ in the two models, but that there exists a mapping between corresponding values of $\alpha$ of the two models. We define corresponding values of $\alpha$ so as to give equal values for $\chibar$. Hence, corresponding values of $\alpha$ and $\Nbar$ yield corresponding states. We assume the existence of a mapping $Q(\alpha_{2})$ such that using $\alpha_{1} = Q(\alpha_2)$ in model 1 yields the same value for $\chibar$ and as that obtained by using $\alpha_2$ in model 2.  Given data for the invariant inverse peak intensity $cN S^{-1}(q^{*})$ from a pair of simulations of these two models with matched values of $\bar{N}$, over a range of values of $\alpha$, it should be possible to collapse this data by plotting the results from model 1 vs. $\alpha_{1}$ and plotting the results from model 2 vs. the corresponding quantity $Q(\alpha_{2})$. 

Because we do not know the mapping function $Q$ {\it a priori}, we must test universality by searching for a 
a mapping function that yields the best possible collapse of results for $cNS^{-1}(q^{*})$ from pairs of simulations with matched values of $\bar{N}$. The required fitting procedure would be meaningless if applied to only a single pair of simulations, with a single value of $\Nbar$: If $cNS^{-1}(q^{*})$ is a monotonic function of $\alpha$, there will always exist a mapping that collapses two such curves, since it will always be possible to identify pairs of values of $\alpha$ that yield the same invariant intensity. The procedure provides a meaningful test, however, if it is applied to several such matched pairs of simulations of the same two models, with a range of values $\Nbar$, by attempting to simultaneously collapse data for several pairs of simulations with different values of $\Nbar$ using a single mapping $Q$ between corresponding values of $\alpha$.

The graphical construction described above requires that we plot the invariant intensity vs. the control parameter $\alpha$ of one of the two models. This choice of coordinate is clearly not unique. If we can find a mapping $Q$ that collapses the data in this representation, then we can also plot the data for model 1 vs. any function $f(\alpha_{1})$ and then plot data for model 2 as a function of a quantity $f(Q(\alpha_{2}))$, without changing the quality of the collapse.

In practice, we have chosen to plot the data for $cNS^{-1}(q^{*})$ for one model in each pair (model 1) as a function of the quantity $N_{1}\chie^{(1)} \equiv z^{\infty}_{1} \alpha N_{1}/\kB T$ used in Fig.~\ref{fig:nomap_wca_bfm} to test the first order perturbation theory for $\chie^{(1)}$, and to plot the data from the other model (model 2) as a function of a quantity $N_{2} \chi_e^{\mathrm{map}}(\alpha_{2})$, in which $\chi_e^{\mathrm{map}}(\alpha_{2})$ is a polynomial whose coefficients are fitted such as to collapse the data. The quantity $N_{2} \chi_e^{\mathrm{map}}$ yields the value of $N_{1} \chie^{(1)}$ in model 1 that corresponds to the same value of $\chibar$, or $\chie N$, as that obtained by using $\alpha_{2}$ in model 2, where $N_{1}$ and $N_{2}$ denote values of $N$ in simulations with matched values of $\bar{N}$. To first order in $\alpha$, $\chi_e^{\mathrm{map}}(\alpha)$ is given by $\chi_e^{\mathrm{map}}(\alpha) \simeq z^{\infty}\alpha/\kB T$, where $z^{\infty}$ is the effective coordination number for model 2. In both of the comparisons shown below, we have fit $\chi_e^{\mathrm{map}}(\alpha_{2})$ using a third polynomial of the form $\chi_e^{\mathrm{map}}(\alpha) = z^{\infty} \alpha + B \alpha^{2} + C\alpha^{3}$ in which the coefficients $B$ and $C$ are adjusted to collapse the data. 

\begin{figure}
\centering\includegraphics[width=\columnwidth]{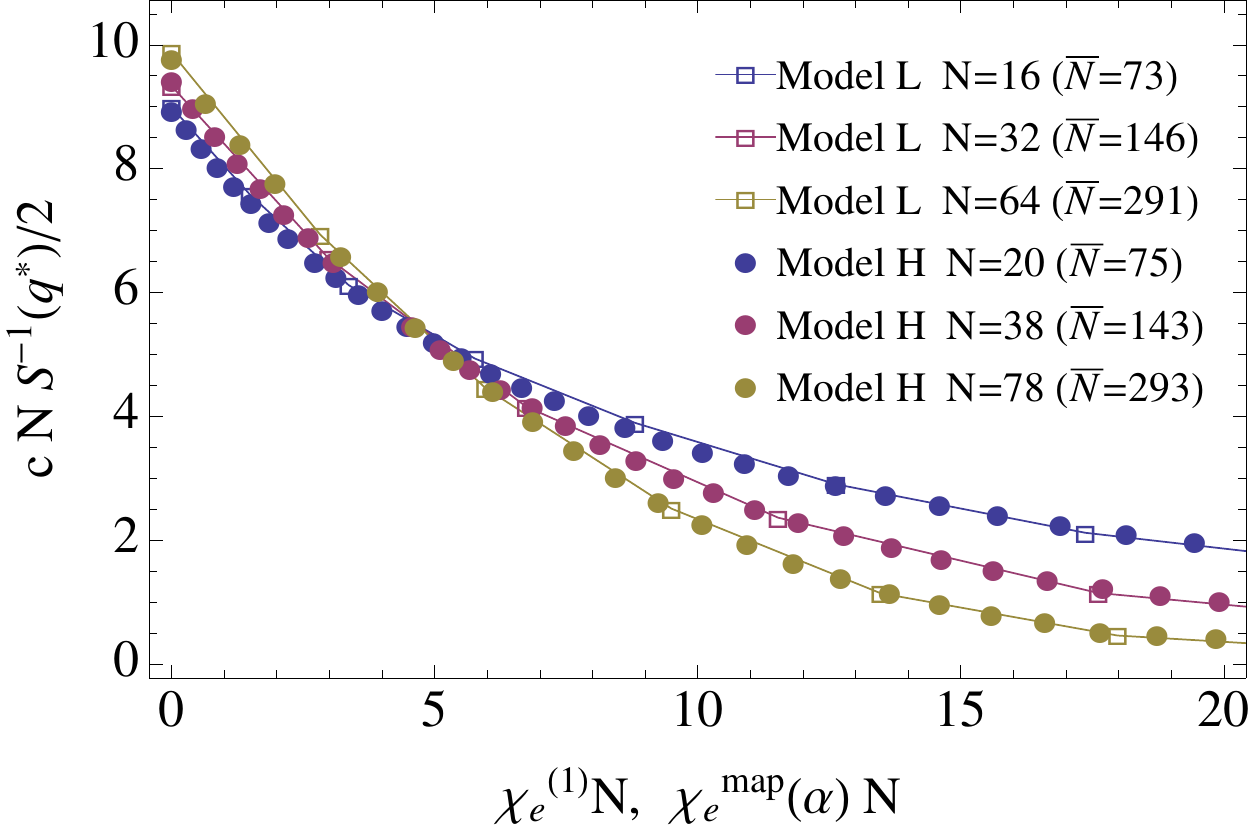}
\caption{
Collapse of the invariant inverse peak intensity $(1/2) cN S^{-1}(q^\star)$ for model L (the lattice model, lines and open squares) and model H (continuum model, solid circles), for three pairs of simulations with nearly matched values of $\Nbar =$ 75, 145 and 292. The data are the same as in Fig.~\ref{fig:nomap_wca_bfm}.  Data for model H are plotted here vs. $\chi_e^{(1)}N= z^{\infty}\alpha N/\kB T$, where $z^{\infty} = 0.2965$. Data for model L are plotted vs. $\chi_e^{\mathrm{map}}(\alpha) N$, using a cubic polynomial $\chi_e^{\mathrm{map}} = z^\infty_{\mathrm{L}}\alpha + 4.94 \alpha^2 + 21.46 \alpha^3$ in which the coefficients of the quadratic and cubic terms have been adjusted to collapse the data.}
\label{fig:map_wca_bfm}
\end{figure}

Figure \ref{fig:map_wca_bfm} shows the results of such an attempt to collapse the peak intensity data for models L (open squares) and H (open circles). The values of $N$ and $\Nbar$ are the same as those used in Fig.~\ref{fig:nomap_wca_bfm}. Here, data for the continuum model is plotted as a function of $\chie(\alpha)N$ while using the first-order approximation $\chie(\alpha) = z^{\infty}\alpha/\kB T$ for $\chie(\alpha)$. Data for the lattice model is plotted vs. $N\chi_e^{\mathrm{map}}(\alpha)$, using a third order polynomial fit for $\chi_e^{\mathrm{map}}(\alpha)$, as described above.  In contrast to the comparison shown in Fig.~\ref{fig:nomap_wca_bfm}, the collapse is nearly perfect. This result confirms that the results of these two models can be adequately described by a single scaling function, but that it is essential to allow for a nonlinear relationship between corresponding values of $\alpha$ in different models.

\begin{figure}[t]
\centering\includegraphics[width=\columnwidth]{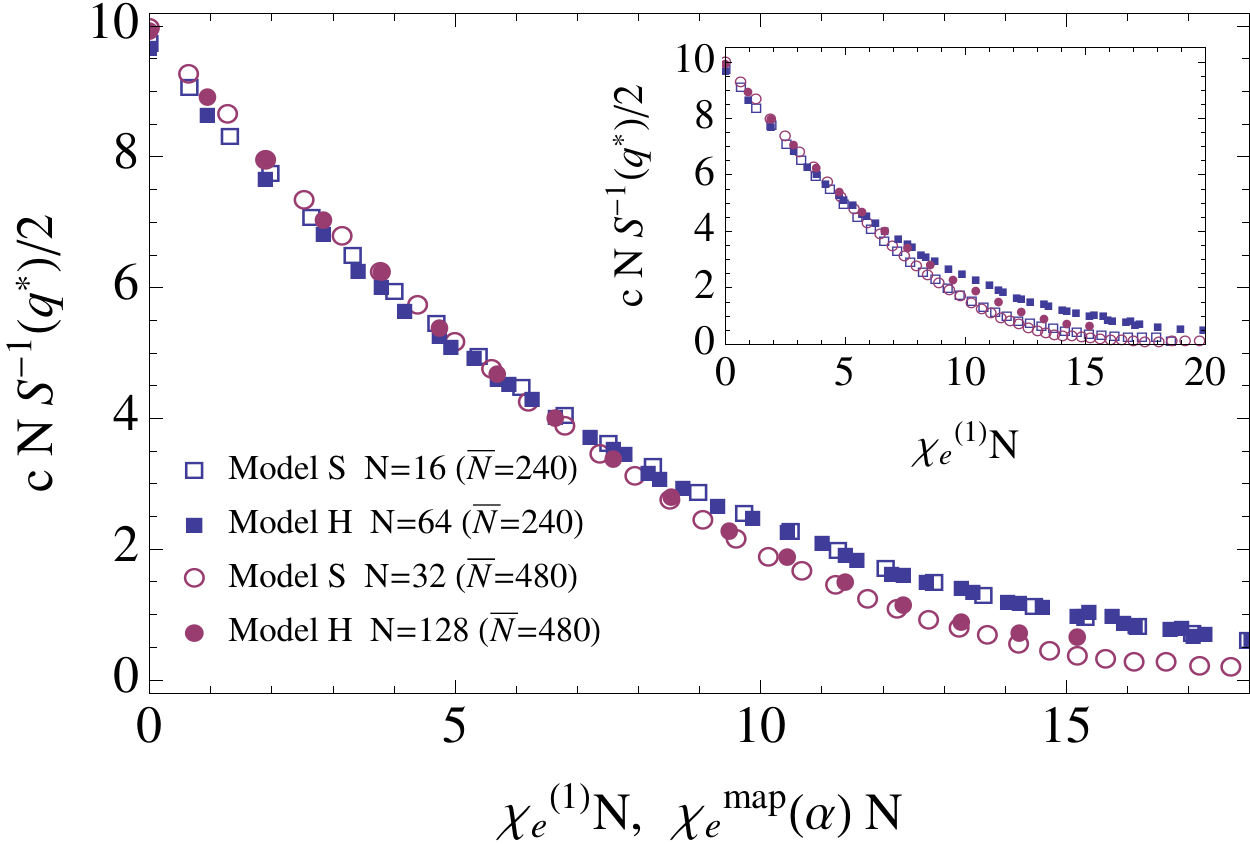}
\caption{Collapse of the invariant inverse structure factor $c N S^{-1}(\qs)$ for model H ('hard') and model S ('soft'), for the same set of simulations as those shown in Fig.~\ref{fig:wcadpd}. The inset shows results for both models plotted using the first-order approximation $\chie^{(1)} (\alpha) = z^\infty \alpha/\kB T$.  In the main plot, data from model H is plotted vs. the first-order approximation for $\chie^{(1)}(\alpha)N$ and data for model S is plotted vs. $\chi_e^{\mathrm{map}}(\alpha) N$, where $\chi_e^{\mathrm{map}}(\alpha)$ is a third order polynomial $\chi_e^{\mathrm{map}}(\alpha)=z^{\infty}_{\mathrm{S}} \alpha + 0.01 \alpha^2 + 0.03 \alpha^3$ in which two coefficients have been adjusted so as to collapse the data. }
\label{fig:map_wca_dpd}
\end{figure}

An analogous comparison of results for models H and S is shown in
Fig. \ref{fig:map_wca_dpd}, for two pairs of simulations with matched
values of $\Nbar = 240$ and $\Nbar = 480$, as in
Fig.~\ref{fig:wcadpd}. Here, the inset shows a comparison in which
data from models is plotted using the first order perturbation theory
estimates for $\chie(\alpha)$, and the main plot shows an attempt to
collapse the data by allowing for a nonlinear mapping. In the main
plot, data for model H is plotted using the first order
perturbation theory for $\chie(\alpha)$, and data for model S
is plotted vs. $N\chi_e^{\mathrm{map}}(\alpha)$, were $\chi_e^{\mathrm{map}}(\alpha)$ is a third order
polynomial with two coefficients that are chosen to collapse the
data. The difference between the models in the inset is less dramatic
than for models H and L. Allowing for a nonlinear mapping again allows
us to obtain nearly perfect collapse of data from two models.

\section{Conclusion}
Theories that are based on the highly idealized standard model of polymer liquids implicitly assume the validity of a principle of corresponding states: They presume that equivalent behavior should be obtained from simulations and experimental systems that can differ in many microscopic details, as long as we compare systems with equal values for all of the dimensionless parameters that appear in the standard model.  The work presented here is motivated by the observation that this is a testable scaling hypothesis, and which can be tested without reference to any quantitative theory, by comparing results of different simulation models. In this paper, we have tested this scaling hypothesis by comparing results for the structure function $S(q)$ from three substantially different simulation models of symmetric diblock polymer melts.  The results strongly support the scaling hypothesis.

The main conceptual difficulty that had to be overcome was the fact that the hypothesis allows all of the quantities of interest to exhibit an unknown dependence on a phenomenological interaction parameter that is an unknown, model-dependent function of the parameters that one can control in a simulation. We devised two tests of universality that did not require knowledge of this dependence. The first of these involved a comparison of the relationship between two observable quantities, the peak wavenumber and the peak intensity. The second involved a test of whether the data for peak intensity from different models could be collapsed by searching for an $N$-independent mapping between corresponding values of the control parameters in different models.  We believe that analogous methods of analyzing data may be useful for analysis of experimental data. In order to compare to experiments, however, it may be important useful to allow for the effects of various types of asymmetry that are not present in these simulations. 

Our results strongly support the scaling hypothesis. Results for the peak position and peak intensity from three substantially different models were found to exhibit almost perfect universality. For comparisons involving model S, this was found even for surprisingly short chains, with as few as $16$ monomers. Notably, the level of agreement between different models for the relationship between the peak wavenumber and peak intensity was substantially better than the already rather good agreement obtained between results of model H and the ROL theory \cite{Qin2012}. This is consistent with the idea that the accuracy of the scaling hypothesis is a measure of the accuracy of the standard model itself, which should have a wider range of validity than the one-loop approximation for this model. 

We emphasize that the goal of this paper was not to test the accuracy of a particular theory for $S(q)$ or other equilibrium properties. Nothing in the analysis presented here referred to a specific theory. Instead, we sought to test whether it was possible for {\it any} coarse-grained theory based on the standard model to describe the data from different simulation models. One consequence of this approach is that our analysis did not yield estimates the interaction parameter $\chi_e$ for any of the models that we have studied. The only way to obtain an estimate of $\chi_e$ for a particular model or experimental system is to compare simulation or experimental results to a quantitative theory. The quality of the resulting estimate can, of course, be no better than the quality of the theory upon which it is based. The analysis presented here shows only that it is possible to construct a theory (or simply a set of empirical correlations) that can describe a variety of simulation models. 

The scaling hypothesis that we discuss here can be easily generalized to ordered phases of block copolymer melts, and to other types of dense polymer liquids, such as polymer mixtures. In future work, we plan to extend this analysis so as to test the universality of behavior in the vicinity of the order-disorder transition of diblock copolymer melts, and in the ordered phases. 

The hypothesis that we tested here is intentionally very simple, in that it allows physical properties to depend only on the variables that appear explicitly in the standard model. It neglects a variety of phenomena that could lead to non-universal deviations from scaling in more realistic models and in experimental systems. Among these are the possible existence of a square-gradient corrections to the local excess free energy, leading to a non-universal $q^{2}$ contribution to $\chi_{e}$, and the possibility of special corrections to the excess free energy arising from chain ends and junctions, as discussed in Ref.~\cite{Grzywacz2007}. The fact that the scaling hypothesis holds to very high accuracy in the models studied here implies, however, that such non-universal corrections must be very small in this particular set of models.  This is encouraging, but it remains to be seen how large may be the deviations from scaling in other simulation models, and in experiments. The type of analysis presented here is, however, the natural starting point for quantifying any such deviations.

\begin{acknowledgments}
  This work was supported by the National Science Foundation (NSF),
  award DMR-0907338, by a postdoctoral fellowship for J.G. from the 
  Deutsche Forschungsgemeinschaft (DFG), GL 733/1-1, and by DFG SFB 602. 
  This research used resources of the Keeneland Computing Facility at the 
  Georgia Institute of Technology, which is supported by the National Science
  Foundation under Contract OCI-0910735, and resources of the Minnesota
  Supercomputing Institute.
\end{acknowledgments}




%

\begin{thebibliography}{33}%
\makeatletter
\providecommand \@ifxundefined [1]{%
 \@ifx{#1\undefined}
}%
\providecommand \@ifnum [1]{%
 \ifnum #1\expandafter \@firstoftwo
 \else \expandafter \@secondoftwo
 \fi
}%
\providecommand \@ifx [1]{%
 \ifx #1\expandafter \@firstoftwo
 \else \expandafter \@secondoftwo
 \fi
}%
\providecommand \natexlab [1]{#1}%
\providecommand \enquote  [1]{``#1''}%
\providecommand \bibnamefont  [1]{#1}%
\providecommand \bibfnamefont [1]{#1}%
\providecommand \citenamefont [1]{#1}%
\providecommand \href@noop [0]{\@secondoftwo}%
\providecommand \href [0]{\begingroup \@sanitize@url \@href}%
\providecommand \@href[1]{\@@startlink{#1}\@@href}%
\providecommand \@@href[1]{\endgroup#1\@@endlink}%
\providecommand \@sanitize@url [0]{\catcode `\\12\catcode `\$12\catcode
  `\&12\catcode `\#12\catcode `\^12\catcode `\_12\catcode `\%12\relax}%
\providecommand \@@startlink[1]{}%
\providecommand \@@endlink[0]{}%
\providecommand \url  [0]{\begingroup\@sanitize@url \@url }%
\providecommand \@url [1]{\endgroup\@href {#1}{\urlprefix }}%
\providecommand \urlprefix  [0]{URL }%
\providecommand \Eprint [0]{\href }%
\providecommand \doibase [0]{http://dx.doi.org/}%
\providecommand \selectlanguage [0]{\@gobble}%
\providecommand \bibinfo  [0]{\@secondoftwo}%
\providecommand \bibfield  [0]{\@secondoftwo}%
\providecommand \translation [1]{[#1]}%
\providecommand \BibitemOpen [0]{}%
\providecommand \bibitemStop [0]{}%
\providecommand \bibitemNoStop [0]{.\EOS\space}%
\providecommand \EOS [0]{\spacefactor3000\relax}%
\providecommand \BibitemShut  [1]{\csname bibitem#1\endcsname}%
\let\auto@bib@innerbib\@empty
\bibitem [{\citenamefont {de~Gennes}(1979)}]{DeGennes1979}%
  \BibitemOpen
  \bibfield  {author} {\bibinfo {author} {\bibfnamefont {P.~G.}\ \bibnamefont
  {de~Gennes}},\ }\href@noop {} {\emph {\bibinfo {title} {{Scaling Concepts in
  Polymer Physics}}}}\ (\bibinfo  {publisher} {Cornell University Press},\
  \bibinfo {address} {Ithaca and London},\ \bibinfo {year} {1979})\BibitemShut
  {NoStop}%
\bibitem [{\citenamefont {des Cloizeaux}\ and\ \citenamefont
  {Jannink}(1990)}]{DesCloizeaux1990}%
  \BibitemOpen
  \bibfield  {author} {\bibinfo {author} {\bibfnamefont {J.}~\bibnamefont {des
  Cloizeaux}}\ and\ \bibinfo {author} {\bibfnamefont {G.}~\bibnamefont
  {Jannink}},\ }\href@noop {} {\emph {\bibinfo {title} {{Polymers in
  Solution}}}}\ (\bibinfo  {publisher} {Clarendon Press},\ \bibinfo {address}
  {Oxford},\ \bibinfo {year} {1990})\BibitemShut {NoStop}%
\bibitem [{\citenamefont {Sch\"{a}fer}(1999)}]{Schafer1999}%
  \BibitemOpen
  \bibfield  {author} {\bibinfo {author} {\bibfnamefont {L.}~\bibnamefont
  {Sch\"{a}fer}},\ }\href@noop {} {\emph {\bibinfo {title} {{Excluded Volume
  Effects in Polymer Solutions}}}}\ (\bibinfo  {publisher} {Springer},\
  \bibinfo {year} {1999})\BibitemShut {NoStop}%
\bibitem [{\citenamefont {Edwards}(1965)}]{Edwards1965}%
  \BibitemOpen
  \bibfield  {author} {\bibinfo {author} {\bibfnamefont {S.~F.}\ \bibnamefont
  {Edwards}},\ }\href {\doibase 10.1088/0370-1328/85/4/301} {\bibfield
  {journal} {\bibinfo  {journal} {Proceedings of the Physical Society}\
  }\textbf {\bibinfo {volume} {85}},\ \bibinfo {pages} {613} (\bibinfo {year}
  {1965})}\BibitemShut {NoStop}%
\bibitem [{\citenamefont {Noda}\ \emph {et~al.}(1984)\citenamefont {Noda},
  \citenamefont {Higo}, \citenamefont {Ueno},\ and\ \citenamefont
  {Fujimoto}}]{Noda1984}%
  \BibitemOpen
  \bibfield  {author} {\bibinfo {author} {\bibfnamefont {I.}~\bibnamefont
  {Noda}}, \bibinfo {author} {\bibfnamefont {Y.}~\bibnamefont {Higo}}, \bibinfo
  {author} {\bibfnamefont {N.}~\bibnamefont {Ueno}}, \ and\ \bibinfo {author}
  {\bibfnamefont {T.}~\bibnamefont {Fujimoto}},\ }\href {\doibase
  10.1021/ma00135a013} {\bibfield  {journal} {\bibinfo  {journal}
  {Macromolecules}\ }\textbf {\bibinfo {volume} {17}},\ \bibinfo {pages} {1055}
  (\bibinfo {year} {1984})}\BibitemShut {NoStop}%
\bibitem [{\citenamefont {Takahashi}\ \emph {et~al.}(1985)\citenamefont
  {Takahashi}, \citenamefont {Isono},\ and\ \citenamefont
  {Noda}}]{Takahashi1985}%
  \BibitemOpen
  \bibfield  {author} {\bibinfo {author} {\bibfnamefont {Y.}~\bibnamefont
  {Takahashi}}, \bibinfo {author} {\bibfnamefont {Y.}~\bibnamefont {Isono}}, \
  and\ \bibinfo {author} {\bibfnamefont {I.}~\bibnamefont {Noda}},\ }\href@noop
  {} {\bibfield  {journal} {\bibinfo  {journal} {Macromolecules}\ }\textbf
  {\bibinfo {volume} {1008}},\ \bibinfo {pages} {1002} (\bibinfo {year}
  {1985})}\BibitemShut {NoStop}%
\bibitem [{\citenamefont {Matsen}(2002)}]{Matsen2002}%
  \BibitemOpen
  \bibfield  {author} {\bibinfo {author} {\bibfnamefont {M.~W.}\ \bibnamefont
  {Matsen}},\ }\href@noop {} {\bibfield  {journal} {\bibinfo  {journal} {J.
  Phys.: Condens. Matter}\ }\textbf {\bibinfo {volume} {14}},\ \bibinfo {pages}
  {R21} (\bibinfo {year} {2002})}\BibitemShut {NoStop}%
\bibitem [{\citenamefont {Fredrickson}\ and\ \citenamefont
  {Helfand}(1987)}]{Fredrickson1987}%
  \BibitemOpen
  \bibfield  {author} {\bibinfo {author} {\bibfnamefont {G.~H.}\ \bibnamefont
  {Fredrickson}}\ and\ \bibinfo {author} {\bibfnamefont {E.}~\bibnamefont
  {Helfand}},\ }\href {\doibase 10.1063/1.453566} {\bibfield  {journal}
  {\bibinfo  {journal} {The Journal of Chemical Physics}\ }\textbf {\bibinfo
  {volume} {87}},\ \bibinfo {pages} {697} (\bibinfo {year} {1987})}\BibitemShut
  {NoStop}%
\bibitem [{\citenamefont {Holyst}\ and\ \citenamefont
  {Vilgis}(1993)}]{Holyst1993}%
  \BibitemOpen
  \bibfield  {author} {\bibinfo {author} {\bibfnamefont {R.}~\bibnamefont
  {Holyst}}\ and\ \bibinfo {author} {\bibfnamefont {T.}~\bibnamefont
  {Vilgis}},\ }\href@noop {} {\bibfield  {journal} {\bibinfo  {journal} {J.
  Chem. Phys.}\ }\textbf {\bibinfo {volume} {99}},\ \bibinfo {pages} {4835}
  (\bibinfo {year} {1993})}\BibitemShut {NoStop}%
\bibitem [{\citenamefont {Kudlay}\ and\ \citenamefont
  {Stepanow}(2003)}]{Kudlay2003}%
  \BibitemOpen
  \bibfield  {author} {\bibinfo {author} {\bibfnamefont {A.}~\bibnamefont
  {Kudlay}}\ and\ \bibinfo {author} {\bibfnamefont {S.}~\bibnamefont
  {Stepanow}},\ }\href {\doibase 10.1063/1.1541612} {\bibfield  {journal}
  {\bibinfo  {journal} {The Journal of Chemical Physics}\ }\textbf {\bibinfo
  {volume} {118}},\ \bibinfo {pages} {4272} (\bibinfo {year}
  {2003})}\BibitemShut {NoStop}%
\bibitem [{\citenamefont {Wang}(2002)}]{Wang2002a}%
  \BibitemOpen
  \bibfield  {author} {\bibinfo {author} {\bibfnamefont {Z.-G.}\ \bibnamefont
  {Wang}},\ }\href {\doibase 10.1063/1.1481761} {\bibfield  {journal} {\bibinfo
   {journal} {The Journal of Chemical Physics}\ }\textbf {\bibinfo {volume}
  {117}},\ \bibinfo {pages} {481} (\bibinfo {year} {2002})}\BibitemShut
  {NoStop}%
\bibitem [{\citenamefont {Grzywacz}\ \emph {et~al.}(2007)\citenamefont
  {Grzywacz}, \citenamefont {Qin},\ and\ \citenamefont {Morse}}]{Grzywacz2007}%
  \BibitemOpen
  \bibfield  {author} {\bibinfo {author} {\bibfnamefont {P.}~\bibnamefont
  {Grzywacz}}, \bibinfo {author} {\bibfnamefont {J.}~\bibnamefont {Qin}}, \
  and\ \bibinfo {author} {\bibfnamefont {D.}~\bibnamefont {Morse}},\ }\href
  {\doibase 10.1103/PhysRevE.76.061802} {\bibfield  {journal} {\bibinfo
  {journal} {Physical Review E}\ }\textbf {\bibinfo {volume} {76}},\ \bibinfo
  {pages} {061802} (\bibinfo {year} {2007})}\BibitemShut {NoStop}%
\bibitem [{\citenamefont {Beckrich}\ \emph {et~al.}(2007)\citenamefont
  {Beckrich}, \citenamefont {Johner}, \citenamefont {Semenov}, \citenamefont
  {Obukhov}, \citenamefont {Beno\^{\i}t},\ and\ \citenamefont
  {Wittmer}}]{Beckrich2007}%
  \BibitemOpen
  \bibfield  {author} {\bibinfo {author} {\bibfnamefont {P.}~\bibnamefont
  {Beckrich}}, \bibinfo {author} {\bibfnamefont {A.}~\bibnamefont {Johner}},
  \bibinfo {author} {\bibfnamefont {A.~N.}\ \bibnamefont {Semenov}}, \bibinfo
  {author} {\bibfnamefont {S.~P.}\ \bibnamefont {Obukhov}}, \bibinfo {author}
  {\bibfnamefont {H.}~\bibnamefont {Beno\^{\i}t}}, \ and\ \bibinfo {author}
  {\bibfnamefont {J.~P.}\ \bibnamefont {Wittmer}},\ }\href {\doibase
  10.1021/ma0626113} {\bibfield  {journal} {\bibinfo  {journal}
  {Macromolecules}\ }\textbf {\bibinfo {volume} {40}},\ \bibinfo {pages} {3805}
  (\bibinfo {year} {2007})}\BibitemShut {NoStop}%
\bibitem [{\citenamefont {Morse}(2006)}]{Morse2006}%
  \BibitemOpen
  \bibfield  {author} {\bibinfo {author} {\bibfnamefont {D.}~\bibnamefont
  {Morse}},\ }\href {\doibase 10.1016/j.aop.2006.02.015} {\bibfield  {journal}
  {\bibinfo  {journal} {Annals of Physics}\ }\textbf {\bibinfo {volume}
  {321}},\ \bibinfo {pages} {2318} (\bibinfo {year} {2006})}\BibitemShut
  {NoStop}%
\bibitem [{\citenamefont {Morse}\ and\ \citenamefont
  {Chung}(2009)}]{Morse2009}%
  \BibitemOpen
  \bibfield  {author} {\bibinfo {author} {\bibfnamefont {D.~C.}\ \bibnamefont
  {Morse}}\ and\ \bibinfo {author} {\bibfnamefont {J.~K.}\ \bibnamefont
  {Chung}},\ }\href {\doibase 10.1063/1.3108460} {\bibfield  {journal}
  {\bibinfo  {journal} {The Journal of Chemical Physics}\ }\textbf {\bibinfo
  {volume} {130}},\ \bibinfo {pages} {224901} (\bibinfo {year}
  {2009})}\BibitemShut {NoStop}%
\bibitem [{\citenamefont {Morse}\ and\ \citenamefont {Qin}(2011)}]{Morse2011}%
  \BibitemOpen
  \bibfield  {author} {\bibinfo {author} {\bibfnamefont {D.~C.}\ \bibnamefont
  {Morse}}\ and\ \bibinfo {author} {\bibfnamefont {J.}~\bibnamefont {Qin}},\
  }\href {\doibase 10.1063/1.3548888} {\bibfield  {journal} {\bibinfo
  {journal} {The Journal of Chemical Physics}\ }\textbf {\bibinfo {volume}
  {134}},\ \bibinfo {pages} {084902} (\bibinfo {year} {2011})}\BibitemShut
  {NoStop}%
\bibitem [{\citenamefont {Qin}\ and\ \citenamefont {Morse}(2012)}]{Qin2012}%
  \BibitemOpen
  \bibfield  {author} {\bibinfo {author} {\bibfnamefont {J.}~\bibnamefont
  {Qin}}\ and\ \bibinfo {author} {\bibfnamefont {D.~C.}\ \bibnamefont
  {Morse}},\ }\href {\doibase 10.1103/PhysRevLett.108.238301} {\bibfield
  {journal} {\bibinfo  {journal} {Phys. Rev. Lett.}\ }\textbf {\bibinfo
  {volume} {238301}},\ \bibinfo {pages} {238301} (\bibinfo {year}
  {2012})}\BibitemShut {NoStop}%
\bibitem [{\citenamefont {Leibler}(1980)}]{Leibler1980}%
  \BibitemOpen
  \bibfield  {author} {\bibinfo {author} {\bibfnamefont {L.}~\bibnamefont
  {Leibler}},\ }\href {\doibase 10.1021/ma60078a047} {\bibfield  {journal}
  {\bibinfo  {journal} {Macromolecules}\ }\textbf {\bibinfo {volume} {13}},\
  \bibinfo {pages} {1602} (\bibinfo {year} {1980})}\BibitemShut {NoStop}%
\bibitem [{\citenamefont {Matsen}\ and\ \citenamefont
  {Schick}(1994)}]{Matsen1994}%
  \BibitemOpen
  \bibfield  {author} {\bibinfo {author} {\bibfnamefont {M.}~\bibnamefont
  {Matsen}}\ and\ \bibinfo {author} {\bibfnamefont {M.}~\bibnamefont
  {Schick}},\ }\href@noop {} {\bibfield  {journal} {\bibinfo  {journal} {Phys.
  Rev. Lett.}\ }\textbf {\bibinfo {volume} {72}},\ \bibinfo {pages} {2660}
  (\bibinfo {year} {1994})}\BibitemShut {NoStop}%
\bibitem [{\citenamefont {Matsen}(1995)}]{Matsen1995}%
  \BibitemOpen
  \bibfield  {author} {\bibinfo {author} {\bibfnamefont {M.~W.}\ \bibnamefont
  {Matsen}},\ }\href {\doibase 10.1021/ma00121a011} {\bibfield  {journal}
  {\bibinfo  {journal} {Macromolecules}\ }\textbf {\bibinfo {volume} {28}},\
  \bibinfo {pages} {5765} (\bibinfo {year} {1995})}\BibitemShut {NoStop}%
\bibitem [{\citenamefont {Tyler}\ and\ \citenamefont
  {Morse}(2005)}]{Tyler2005}%
  \BibitemOpen
  \bibfield  {author} {\bibinfo {author} {\bibfnamefont {C.}~\bibnamefont
  {Tyler}}\ and\ \bibinfo {author} {\bibfnamefont {D.}~\bibnamefont {Morse}},\
  }\href {\doibase 10.1103/PhysRevLett.94.208302} {\bibfield  {journal}
  {\bibinfo  {journal} {Physical Review Letters}\ }\textbf {\bibinfo {volume}
  {94}},\ \bibinfo {pages} {208302} (\bibinfo {year} {2005})}\BibitemShut
  {NoStop}%
\bibitem [{\citenamefont {Carmesin}\ and\ \citenamefont
  {Kremer}(1988)}]{Carmesin1988}%
  \BibitemOpen
  \bibfield  {author} {\bibinfo {author} {\bibfnamefont {I.}~\bibnamefont
  {Carmesin}}\ and\ \bibinfo {author} {\bibfnamefont {K.}~\bibnamefont
  {Kremer}},\ }\href@noop {} {\bibfield  {journal} {\bibinfo  {journal}
  {Macromolecules}\ }\textbf {\bibinfo {volume} {21}},\ \bibinfo {pages} {2819}
  (\bibinfo {year} {1988})}\BibitemShut {NoStop}%
\bibitem [{\citenamefont {M\"{u}ller}\ and\ \citenamefont
  {Binder}(1995)}]{Mueller1995}%
  \BibitemOpen
  \bibfield  {author} {\bibinfo {author} {\bibfnamefont {M.}~\bibnamefont
  {M\"{u}ller}}\ and\ \bibinfo {author} {\bibfnamefont {K.}~\bibnamefont
  {Binder}},\ }\href {\doibase 10.1021/ma00110a016} {\bibfield  {journal}
  {\bibinfo  {journal} {Macromolecules}\ }\textbf {\bibinfo {volume} {28}},\
  \bibinfo {pages} {1825} (\bibinfo {year} {1995})}\BibitemShut {NoStop}%
\bibitem [{\citenamefont {M\"{u}ller}(1995)}]{Mueller1995a}%
  \BibitemOpen
  \bibfield  {author} {\bibinfo {author} {\bibfnamefont {M.}~\bibnamefont
  {M\"{u}ller}},\ }\href@noop {} {\bibfield  {journal} {\bibinfo  {journal}
  {Macromolecules}\ }\textbf {\bibinfo {volume} {28}},\ \bibinfo {pages} {6556}
  (\bibinfo {year} {1995})}\BibitemShut {NoStop}%
\bibitem [{\citenamefont {Kremer}\ and\ \citenamefont
  {Grest}(1990)}]{Kremer1990}%
  \BibitemOpen
  \bibfield  {author} {\bibinfo {author} {\bibfnamefont {K.}~\bibnamefont
  {Kremer}}\ and\ \bibinfo {author} {\bibfnamefont {G.~S.}\ \bibnamefont
  {Grest}},\ }\href {\doibase 10.1063/1.458541} {\bibfield  {journal} {\bibinfo
   {journal} {The Journal of Chemical Physics}\ }\textbf {\bibinfo {volume}
  {92}},\ \bibinfo {pages} {5057} (\bibinfo {year} {1990})}\BibitemShut
  {NoStop}%
\bibitem [{\citenamefont {Groot}\ and\ \citenamefont
  {Madden}(1998)}]{Groot1998}%
  \BibitemOpen
  \bibfield  {author} {\bibinfo {author} {\bibfnamefont {R.~D.}\ \bibnamefont
  {Groot}}\ and\ \bibinfo {author} {\bibfnamefont {T.~J.}\ \bibnamefont
  {Madden}},\ }\href {\doibase 10.1063/1.476300} {\bibfield  {journal}
  {\bibinfo  {journal} {The Journal of Chemical Physics}\ }\textbf {\bibinfo
  {volume} {108}},\ \bibinfo {pages} {8713} (\bibinfo {year}
  {1998})}\BibitemShut {NoStop}%
\bibitem [{\citenamefont {Pike}\ \emph {et~al.}(2009)\citenamefont {Pike},
  \citenamefont {Detcheverry}, \citenamefont {M\"{u}ller},\ and\ \citenamefont
  {de~Pablo}}]{Pike2009}%
  \BibitemOpen
  \bibfield  {author} {\bibinfo {author} {\bibfnamefont {D.~Q.}\ \bibnamefont
  {Pike}}, \bibinfo {author} {\bibfnamefont {F.~a.}\ \bibnamefont
  {Detcheverry}}, \bibinfo {author} {\bibfnamefont {M.}~\bibnamefont
  {M\"{u}ller}}, \ and\ \bibinfo {author} {\bibfnamefont {J.~J.}\ \bibnamefont
  {de~Pablo}},\ }\href {\doibase 10.1063/1.3187936} {\bibfield  {journal}
  {\bibinfo  {journal} {The Journal of chemical physics}\ }\textbf {\bibinfo
  {volume} {131}},\ \bibinfo {pages} {084903} (\bibinfo {year}
  {2009})}\BibitemShut {NoStop}%
\bibitem [{\citenamefont {Wang}\ and\ \citenamefont {Yin}(2009)}]{Wang2009a}%
  \BibitemOpen
  \bibfield  {author} {\bibinfo {author} {\bibfnamefont {Q.}~\bibnamefont
  {Wang}}\ and\ \bibinfo {author} {\bibfnamefont {Y.}~\bibnamefont {Yin}},\
  }\href {\doibase 10.1063/1.3086606} {\bibfield  {journal} {\bibinfo
  {journal} {The Journal of chemical physics}\ }\textbf {\bibinfo {volume}
  {130}},\ \bibinfo {pages} {104903} (\bibinfo {year} {2009})}\BibitemShut
  {NoStop}%
\bibitem [{\citenamefont {Weeks}\ \emph {et~al.}(1971)\citenamefont {Weeks},
  \citenamefont {Chandler},\ and\ \citenamefont {Andersen}}]{Weeks1971}%
  \BibitemOpen
  \bibfield  {author} {\bibinfo {author} {\bibfnamefont {J.~D.}\ \bibnamefont
  {Weeks}}, \bibinfo {author} {\bibfnamefont {D.}~\bibnamefont {Chandler}}, \
  and\ \bibinfo {author} {\bibfnamefont {H.~C.}\ \bibnamefont {Andersen}},\
  }\href@noop {} {\bibfield  {journal} {\bibinfo  {journal} {J. Chem. Phys.}\
  }\textbf {\bibinfo {volume} {54}},\ \bibinfo {pages} {5237} (\bibinfo {year}
  {1971})}\BibitemShut {NoStop}%
\bibitem [{Note1()}]{Note1}%
  \BibitemOpen
  \bibinfo {note} {See https://github.com/dmorse/simpatico}\BibitemShut
  {NoStop}%
\bibitem [{\citenamefont {Anderson}\ \emph {et~al.}(2008)\citenamefont
  {Anderson}, \citenamefont {Lorenz},\ and\ \citenamefont
  {Travesset}}]{Anderson2008}%
  \BibitemOpen
  \bibfield  {author} {\bibinfo {author} {\bibfnamefont {J.}~\bibnamefont
  {Anderson}}, \bibinfo {author} {\bibfnamefont {C.}~\bibnamefont {Lorenz}}, \
  and\ \bibinfo {author} {\bibfnamefont {a.}~\bibnamefont {Travesset}},\ }\href
  {\doibase 10.1016/j.jcp.2008.01.047} {\bibfield  {journal} {\bibinfo
  {journal} {Journal of Computational Physics}\ }\textbf {\bibinfo {volume}
  {227}},\ \bibinfo {pages} {5342} (\bibinfo {year} {2008})}\BibitemShut
  {NoStop}%
\bibitem [{\citenamefont {Wittmer}\ \emph {et~al.}(2007)\citenamefont
  {Wittmer}, \citenamefont {Beckrich}, \citenamefont {Meyer}, \citenamefont
  {Cavallo}, \citenamefont {Johner},\ and\ \citenamefont
  {Baschnagel}}]{Wittmer2007}%
  \BibitemOpen
  \bibfield  {author} {\bibinfo {author} {\bibfnamefont {J.}~\bibnamefont
  {Wittmer}}, \bibinfo {author} {\bibfnamefont {P.}~\bibnamefont {Beckrich}},
  \bibinfo {author} {\bibfnamefont {H.}~\bibnamefont {Meyer}}, \bibinfo
  {author} {\bibfnamefont {a.}~\bibnamefont {Cavallo}}, \bibinfo {author}
  {\bibfnamefont {a.}~\bibnamefont {Johner}}, \ and\ \bibinfo {author}
  {\bibfnamefont {J.}~\bibnamefont {Baschnagel}},\ }\href {\doibase
  10.1103/PhysRevE.76.011803} {\bibfield  {journal} {\bibinfo  {journal}
  {Physical Review E}\ }\textbf {\bibinfo {volume} {76}},\ \bibinfo {pages} {1}
  (\bibinfo {year} {2007})}\BibitemShut {NoStop}%
\bibitem [{\citenamefont {M\"{u}ller}(1999)}]{Muller1999}%
  \BibitemOpen
  \bibfield  {author} {\bibinfo {author} {\bibfnamefont {M.}~\bibnamefont
  {M\"{u}ller}},\ }\href@noop {} {\bibfield  {journal} {\bibinfo  {journal}
  {Macromol. Theory Simul.}\ }\textbf {\bibinfo {volume} {8}},\ \bibinfo
  {pages} {343} (\bibinfo {year} {1999})}\BibitemShut {NoStop}%
\end{thebibliography}%
\end{document}